\documentstyle[aps,prl]{revtex}

\begin{document}

\title{A Comment on ``Dynamics of Weak First Order Phase Transitions''}
\author{Geoffrey~Harris}
\address{Enrico Fermi Institute\\
University of Chicago\\
Chicago, IL 60637, USA}
\author{Gerard~Jungman}
\address{ Physics Department\\
Syracuse University\\
Syracuse, NY 13244, USA }

\date{July 1995}
\maketitle

Nucleation and spinodal decomposition, as mechanisms
for first-order transitions,
have been studied for many years.
A pertinent
issue is whether
or not
there is a sharp distinction between these two processes.
In Ref. \cite{gleiser}, M. Gleiser analyzes a lattice model with
one free
parameter $\alpha$ which is tuned to control the strength of a
first--order
transition.  He argues that there exists a critical value
$\alpha_c$
distinguishing between strong and weak first--order
transitions
and suggests that only for $\alpha > \alpha_c$ does phase transformation
proceed
by nucleation.   We believe that the numerical results in this
work are correct, but will argue against the interpretation of the
theory at $\alpha_c$ as a boundary between weak and strong phase
transitions.

The phenomenology of phase transitions can be described by the
Landau--Ginzburg Hamiltonian
\begin{equation}
\protect\label{lg}
{H \over \theta} = {1 \over \theta}
\left( {1\over 2}(\nabla \phi)^2 \pm m^2 \phi^2
 + {\lambda\over 4} \phi^4
 + \beta (t) \phi +  \ldots . \right).
\end{equation}
As the system is cooled, $\beta (t)$ changes sign,
the minimum of
the free--energy shifts discontinuously, and phase transformation occurs.
In Ref. \cite{gleiser}, the above Hamiltonian,
with $\beta (t) \equiv 0$ and a positive coefficient
of $\phi ^2$, is adopted
as a {\it lattice} Hamiltonian.  This free energy has two degenerate
minima;
$\phi$ is placed throughout the lattice in the left--most minima
and its relaxation (with second order Langevin dynamics) is studied;
$\alpha $ controls the height of the hump between the
two minima.
At a
certain
non--zero value of $\alpha$, the relaxation dynamics
becomes critical.  The size of the potential hump
is interpreted as a measure of the strength of a first--order
transition that would occur under cooling, implemented
by
a
time--dependent parity--odd
perturbation of
the Hamiltonian.  For $\alpha
\gg \alpha_c$, the bubble free energy is large and nucleation should
adequately describe
the relaxation.  It is then inferred that
a different relaxation process may
operate
when
$\alpha < \alpha _c$.

This
description is misleading because it depends on the
shape of the lattice (bare) potential.  The physics of the phase
transformation
is modeled by the {\it coarse}--{\it grained} potential,
related to the bare
potential
by renormalization.
Criticality
of the relaxation process depends only on the long--time equilibrium
behavior.
Thus, we only need the
renormalization corrections
for a classical statistical mechanical model at finite temperature
$\theta$,
or equivalently,
a quantum field theory with $\hbar = 1/\theta$.
Explicitly, consider
the one-loop correction to the quadratic
term, which depends on the
inverse lattice spacing $\Lambda$ as
\begin{equation}
\protect\label{renorm}
	H_{ct} = {1\over 4\pi} 3 \lambda \theta
		\ln \left( {\Lambda^2 \over \mu^2(m,\lambda \theta)} \right)
		\; \phi^2;
\end{equation}
the finite part of the counterterm, $\mu$, is determined by imposing
renormalization conditions.
The leading order $\Lambda $ dependent piece reduces the barrier
in the coarse--grained free energy by an additive correction.
We
anticipate that at some value
of $\alpha$, presumably $\alpha_c$, the quadratic term and bump in the
continuum effective
potential
vanish.  This corresponds to the
second--order
phase transition in the $\phi^4$ system, which is in the
Ising
universality class.

Indeed, with this correspondence in mind, simulations of the lattice
Hamiltonian of Ref. \cite{gleiser}
have been previously performed.
In Ref. \cite{milchev}, Monte Carlo
simulations
are used to extract Ising critical exponents from this
model.
In Ref. \cite{toralchak}, Monte Carlo and Langevin simulations are used
to obtain
the critical line of the lattice $\phi^4$ model.
The coupling $\chi$ of
Ref. \cite{toralchak}
is exactly the quantity $\lambda\theta_c $ of Ref. \cite{gleiser};
the coupling $\theta^{TC}$ of Ref. \cite{toralchak},
which we distinguish by
a
superscript $TC$,
equals $(\theta _c^2 -1)/2$ in Ref. \cite{gleiser},
with $\theta_c = (1 - 2\alpha ^2/9\lambda )^{-1/2}$.
In Ref. \cite{gleiser},
criticality occurs at
$(\lambda, \alpha ) = (.1, .36)$. This corresponds to
$(\chi, \theta^{TC}) = (.119,.202)$, which lies
within statistical error
on the critical line
presented in Table II and Figure 8 of Ref.
\cite{toralchak}.
We have repeated these numerical simulations, using Langevin dynamics
as in Ref. \cite{toralchak}. We find results consistent with those above.
Thus the critical behavior
in Ref. \cite{gleiser} is in the Ising universality class,
with vanishing coarse--grained barrier
energy.
It is not appropriately described in
the continuum limit as a boundary between large and small
barrier energies.

G.H. and G.J. were supported by DOE grant DE-FG02-85ER-40231
at Syracuse University.  Additionally, G.H. was supported by DOE
grant DE-FG02-90ER-40560 at the University of Chicago.


\begin{references}
\bibitem{gleiser}
	M. Gleiser, Phys. Rev. Lett. {\bf 73}, 3495 (1994).
\bibitem{milchev}
A. Milchev, D.W. Heermann and K. Binder, J. Stat. Phys. {\bf 44}, 749 (1986).
\bibitem{toralchak}
R. Toral and A. Chakrabarti, Phys. Rev. {\bf B42}, 2445 (1990).
\end{references}
\end{document}